# Cardiac MRI Image Segmentation for Left Ventricle and Right Ventricle using Deep Learning

Bosung Seo, Daniel Mariano, John Beckfield, Vinay Madenur, Yuming Hu, Tony Reina, Marcus Bobar, Mai H. Nguyen[1], Ilkay Altintas

**Abstract**

The goal of this project is to use magnetic resonance imaging (MRI) data to provide an end-to-end analytics pipeline for left and right ventricle (LV and RV) segmentation. Another aim of the project is to find a model that would be generalizable across medical imaging datasets. We utilized a variety of models, datasets, and tests to determine which one is well suited to this purpose. Specifically, we implemented three models (2-D U-Net, 3-D U-Net, and DenseNet), and evaluated them on four datasets (Automated Cardiac Diagnosis Challenge, MICCAI 2009 LV, Sunnybrook Cardiac Data, MICCAI 2012 RV). While maintaining a consistent preprocessing strategy, we tested the performance of each model when trained on data from the same dataset as the test data, and when trained on data from a different dataset than the test dataset. Data augmentation was also used to increase the adaptability of the models. The results were compared to determine performance and generalizability.

**Introduction**

Currently, the process of manually analyzing cardiac MRI images is time-consuming and labor-intensive. There is no easy way to segment the LV which is the largest chamber in the heart and plays a critical role in cardiac function. LV segmentation refers to the task of detecting the LV contour, especially of the endocardial surface. LV segmentation using MRI images is a challenging task as difficulties arise from many sources. The image quality of different parts of the heart varies. The middle of the heart is generally clear, but the top and bottom slices can be very hard to determine the correct LV contour, even for an expert. Other parts of the heart may obscure the desired section of heart. Even for an expert, it can take up to an hour to manually segment the LV for a single patient. There will always be discrepancies between the various human labelers.

The RV segmentation is challenging due to the fuzziness of the cavity borders due to blood flow and partial volume effect, the presence of trabeculations (wall irregularities) in the cavity, which have the same grey level as the surrounding myocardium, and the complex crescent shape of the RV, which varies according to the imaging slice level. As a consequence, RV functional assessment has long been considered secondary compared to that of the LV, leaving the problem of RV segmentation wide open.

As such, we would like to see how deep learning can be used to automate some of the steps in this process to provide faster and more consistent image analysis results. Our inputs will be images from the ACDC,

---

[1] Corresponding author: Mai H. Nguyen, mhnguyen@ucsd.edu

Sunnybrook, and MICCAI RV, LV-2011 datasets in the DICOM and NIfTI file formats. One thing to note is that only the images with contours will be normalized as that is what we will be able to train and verify as predictions against. Our targets will be Tensor arrays with normalized sizes and contrasts. Our preprocessing will read in the images, convert them to Numpy arrays, and then normalize them by size, followed by contrast. Last, the Numpy arrays are converted into Tensor arrays, which are ready to use with Keras. This normalization method will be implemented on the images and contours. The normalized images will be used in the training and testing of the network model.

There are many variations that can happen from image to image, and from dataset to dataset. There are different strengths and types of MRI machines. Radiologists have different skill levels. There is a lot of variation from patient to patient. An MRI image of a patient with a suspected left ventricle problem will be centered differently than a patient with a suspected right ventricle problem. Even how well patients are able to hold their breath will affect MRI quality. Due to the extreme variability, a model that performs very well on one dataset might not perform well on another. We will be testing on multiple datasets to try to find the models and preprocessing steps that best generalize to a high number of different datasets.

**Datasets**

Each MRI image has been taken on a different patient with a different MRI machine by a different radiologist. Due to these factors, the MRI images across each dataset are not consistent from one to the next. Across the datasets, the image's pixel spacing, image size, and image orientation are different which is something that we had to consider when figuring out what preprocessing steps we wanted to take.

For our project purposes, the NIfTI images from the ACDC needed to be converted to a similar format as the DICOM Images. This was accomplished by breaking out the NIfTI images by the slice. These frames within the slice are then converted into 2-D numpy arrays to be used for preprocessing. The DICOM images are converted into numpy arrays for preprocessing as well.

In the below figure, is a summary of the properties of the datasets that we used:

| Dataset Name | Number of Patients | Total Number of Training Images | Number of Training Images with labels | Number of Slices per Patient | Number of Frames per Slice with labels | Size of Normalized Images (MB) - Khened | Size of Normalized Images (MB) - Isensee |
|---|---|---|---|---|---|---|---|
| SB | 45 | 51545 | 420 | 5 - 6 | 1 - 2 | 200 | 150 |
| ACDC | 100 | 25351 | 1902 | 6 - 18 | 2 | 470 | 715 |
| MICCAI RV | 16 | 3940 | 243 | 6 - 11 | 1 - 2 | 115 | 100 |

| | | | | | | |
|---|---|---|---|---|---|---|
| LV-2011 | 100 | 29086 | 2522 | 8 - 24 | 2 | 1230 | 1024 |

Figure 1: Datasets Summary

**ACDC**

The Automated Cardiac Diagnosis Challenge(ACDC) dataset consists of Short Axis (SAX) view MRI Images for 100 patients (3.3 GB) in the NIfTI (Neuroimaging Informatics Technology Initiative) image format. Each patient directory consists of 4-D NIfTI Format Images. Contour files have been provided for the End-Systolic and End-Diastolic images for each patient. These contours were drawn to follow the limit defined by the aortic valve. The expert references are manually-drawn 3D volumes of the LV and RV cavities as well as the myocardium, both at the ED and ES slices. The acquisitions were obtained over a 6 year period using two MRI scanners of different magnetic strengths (1.5 T (Siemens Area, Siemens Medical Solutions, Germany) and 3.0 T (Siemens Trio Tim, Siemens Medical Solutions, Germany)). Cine MR images were acquired in breath hold with a retrospective or prospective gating and with a SSFP sequence in short axis orientation. Particularly, a series of short axis slices cover the LV from the base to the apex, with a thickness of 5 mm (or sometimes 8 mm) and sometimes an interslice gap of 5 mm (then one image every 5 or 10 mm, according to the examination). The spatial resolution goes from 1.37 to 1.68 mm2/pixel and 28 to 40 images cover completely or partially the cardiac cycle (in the second case, with prospective gating, only 5 to 10 % of the end of the cardiac cycle was omitted), all depending on the patient.

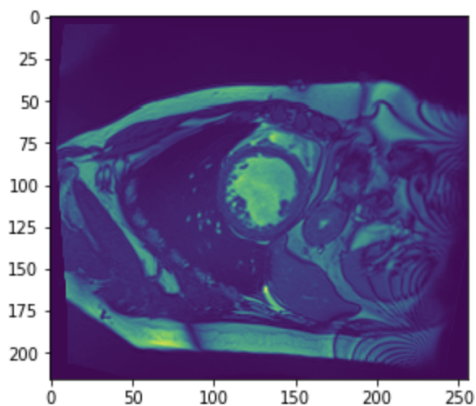

Figure 2: An example raw ACDC image

**MICCAI RV**

The Right Ventricle Segmentation Dataset MICCAI 2012 dataset contains 48 cardiac cine-MR data (1.5 GB) with contours drawn by one cardiac radiologist (16 for training, 32 for testing). The images have been zoomed and cropped to a 256x216 (or 216x256) pixel ROI, leaving the LV visible for joint ventricle segmentation, as necessary. The contours consist of delineated endocardial and epicardial borders of the RV on short axis slices at ED and ES. Trabeculae and papillary muscles are also included in the ventricular cavity. Cardiac MR examinations were performed at 1.5T (Symphony Tim®, Siemens Medical Systems, Erlangen, Germany). A dedicated eight-element phased-array cardiac coil was used. Retrospectively synchronized balanced steady-state free precession sequences were performed for cine analysis, with repeated breath-holds of 10-15 s. All conventional planes (2-, 3- and 4-chamber views) were acquired and a total of 8-12 contiguous cine short axis slices were performed from the base to the apex of the ventricles. Sequence parameters were as follows: TR = 50 ms; TE = 1.7 ms; flip angle = 55°; slice thickness = 7 mm;

matrix size = 256 x 216; Field of view = 360-420 mm; 20 images per cardiac cycle. Clinical indications were represented by a panel of the currently most frequent cardiac MRI indications in patients with AHD: myocarditis, ischaemic cardiomyopathy, suspicion of arrhythmogenic right ventricular dysplasia, dilated cardiomyopathy, hypertrophic cardiomyopathy, aortic stenosis.

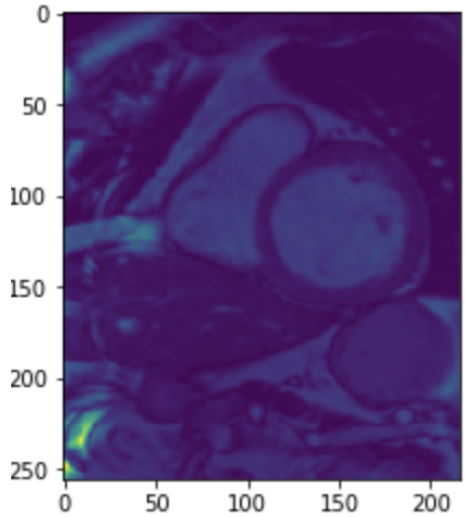
*Figure 3: An example raw MICCAI RV image*

**Sunnybrook**

The Sunnybrook Cardiac Data(SCD) dataset consists of 45 cardiac cine-MRI images (1.6 GB) from a mixed group of patients and pathologies: healthy, hypertrophy, heart failure with infarction, and heart failure without infarction. The MRI images are in the DICOM (Digital Imaging and Communications in Medicine) image format that consists of several metadata parameters about the patient and the image. For each patient record, there is a set of hand drawn contours (one for the endocardium and one for the epicardium) for End Diastolic (ED) and End Systolic (ES) slices. The contours were drawn by Perry Radau from the Sunnybrook Health Science Centre. The contours are available in text files that consist of the contour points. Cine steady state free precession (SSFP) MR short axis (SAX) images were obtained with a 1.5T GE Signa MRI. All the images were obtained during 10-15 second breath-holds with a temporal resolution of 20 cardiac phases over the heart cycle, and scanned from the ED phase. Six to 12 SAX images were obtained from the atrioventricular ring to the apex (thickness=8mm, gap=8mm, FOV=320mm320mm, matrix= 256256).

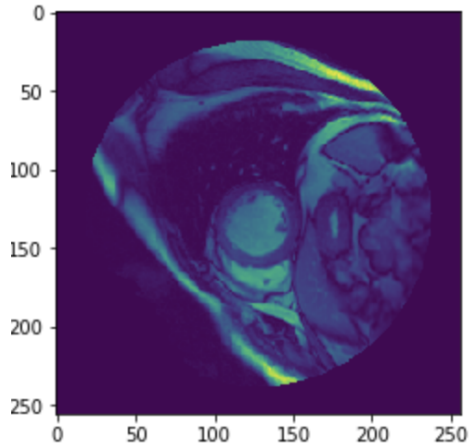
*Figure 4: An example raw Sunnybrook image*

**LV**

The 2011 LV Segmentation Challenge Dataset dataset is of one hundred patients randomly selected from the DETERMINE cohort (Defibrillators To Reduce Risk by Magnetic Resonance Imaging Evaluation) (Kadish et al., 2009). DETERMINE is a prospective, multicenter, randomized clinical trials in patients with coronary artery diseases and mild-to-moderate left ventricular dysfunction. The primary objective of the study is to test the hypothesis that implantable cardioverter defibrillator (ICD) therapy in combination with medical therapy, in patients with myocardial infarct greater than or equal to 15% of the left heart muscle mass (as measured by CMR), improves long term survival compared to medical therapy alone. One hundred were made available as training data, with manual segmentation, and the other hundred were reserved for validation. The DETERMINE study comprises of patients with coronary artery disease and regional wall motion abnormalities due to prior myocardial infarction. This is a clinically important patient group since mass and volume are important diagnostic and prognostic indicators of adverse remodeling. Studies were acquired at multiple sites using multiple scanner vendors. The data were made available through the Cardiac Atlas Project (Fonseca et al., 2011). The CMR images were based on the steady-state free precession (SSFP) pulse sequence. CMR parameters varied between cases giving a heterogeneous mix of scanner types and imaging parameters. MR scanner systems were GE Medical Systems (Signa 1.5T), Philips Medical Systems (Achieva 1.5T, 3.0T, and Intera 1.5T), and Siemens (Avanto 1.5T, Espree 1.5T and Symphony 1.5T). Typical short-axis slice parameters were either a 6 mm slice thickness with 4 mm gap or 8 mm slice thickness with 2 mm gap. Image size was ranging from 138 × 192 to 512 × 512 pixels. The temporal resolution was between 19 and 30 frames. Long axis images in the four and two chamber orientations were also available.

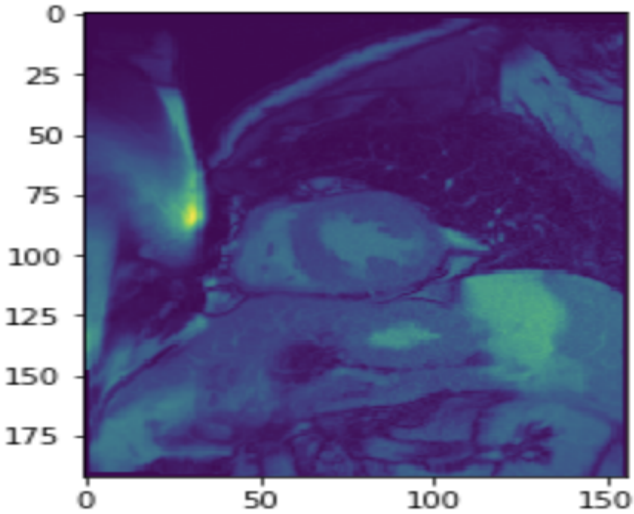
*Figure 5: An example raw LV-2011 image*

**Data Pipeline**

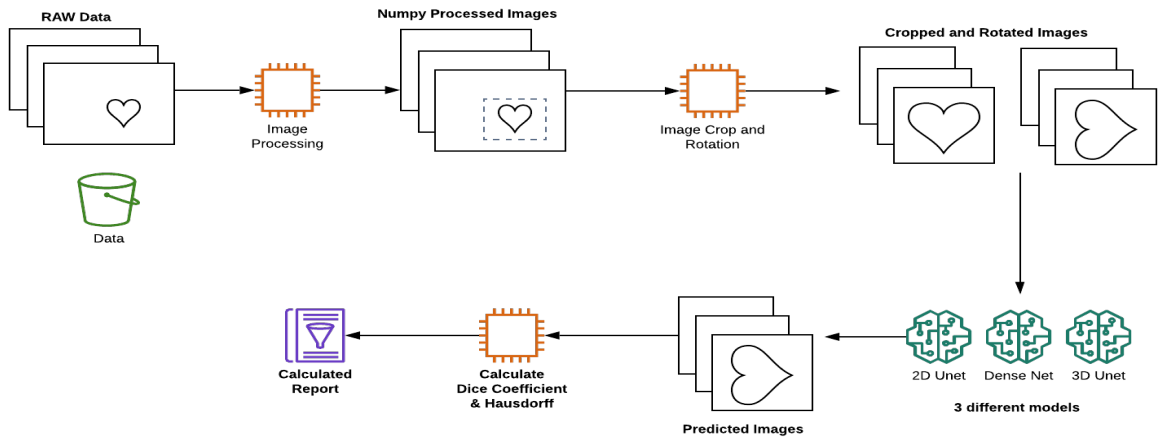
*Figure 6: Data Pipeline*

The above figure illustrates the prediction pipeline. The raw images (dicom and nifti format) are first converted to numpy arrays. Then a bunch of preprocessing steps like cropping and contrast normalization are performed on the data. The processed data is then fed into the various trained models to generate the predictions. Finally using the predictions and ground truth, the various metrics like Dice coefficient & Hausdorff value are calculated and uploaded to AWS S3.

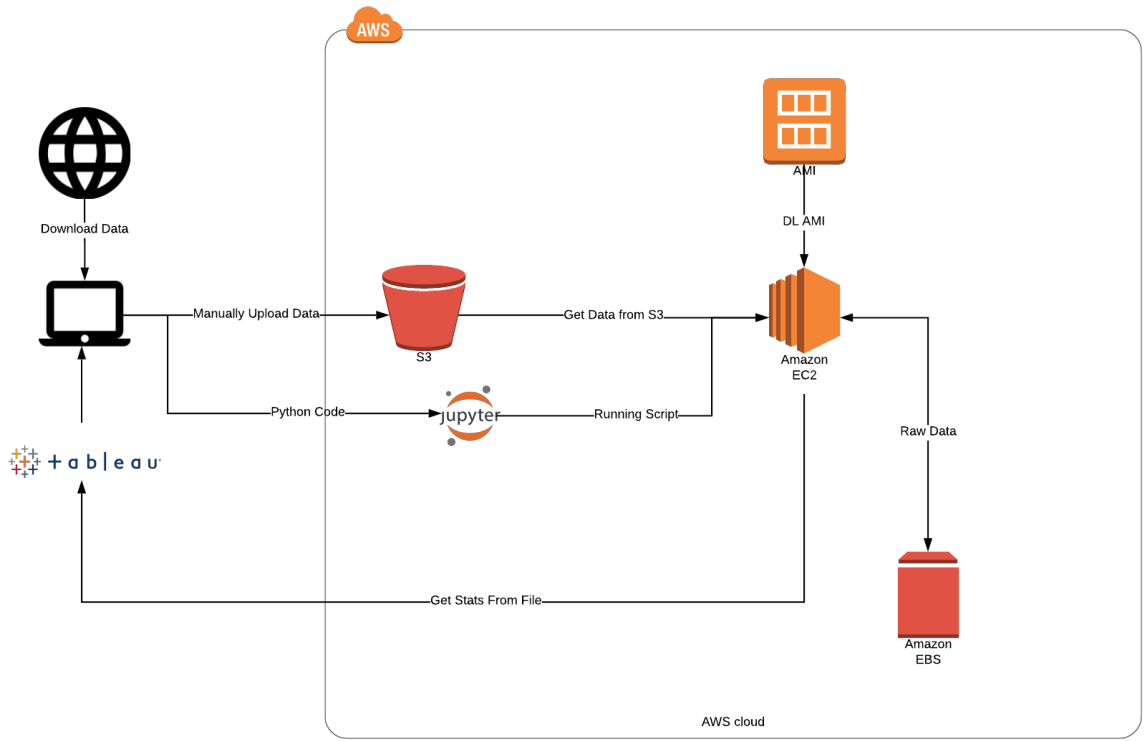

*Figure 7: Infrastructure Pipeline*

The training of all the models was done on AWS. The AWS Deep Learning AMI was used to launch EC2 instances with GPU (e.g. p3.2xlarge & p3.8xlarge). In order to leverage the advantages of the GPU you need to satisfy some criteria:
1. You need to have GPU drivers setup correctly.
2. You need to have libraries that can leverage all the GPU power during the training of the neural network. The libraries need to be compatible with the drivers.
3. You need to have neural network framework that have been compiled with the libraries that you have.

AWS Deep Learning AMI is an Amazon Machine Image which comes pre-configured with NVIDIA drivers, as well as the latest releases of the most popular deep learning frameworks like Keras, Tensorflow etc. All the data was stored on AWS S3 and was copied to the EBS volumes attached to the EC2 instances before training of the models started. The final results were copied back to S3 and were used in Tableau visualizations.

**Preprocessing Experiments**

During the course of our project, we first attempted using the same preprocessing steps as in previous work [4]. If the image is from the ACDC dataset, it will be flipped by 180 degrees to match the orientation of other datasets. Next, the image will be rescaled based on the image's pixel spacing so we get each pixel representing the same amount of area, 1mm x 1mm. After the image is rescaled, it will be resized/cropped from the center of the image to be 176 x 176 pixels. Then, Contrast Limited Adaptive Histogram

Equalization (CLAHE) will be applied to each 1 x 1 tile of the image. This normalizes the contrast of the image. Last, the pixel intensity of the image will be normalized using Min-Max normalization.

In order to assess how the images were altered from their original form and if similar results were produced, viewing the images is the best way to identify the changes. In the two figures below, one can see two images from two different datasets where normalization was applied. Each stage of the preprocessing steps are shown below.

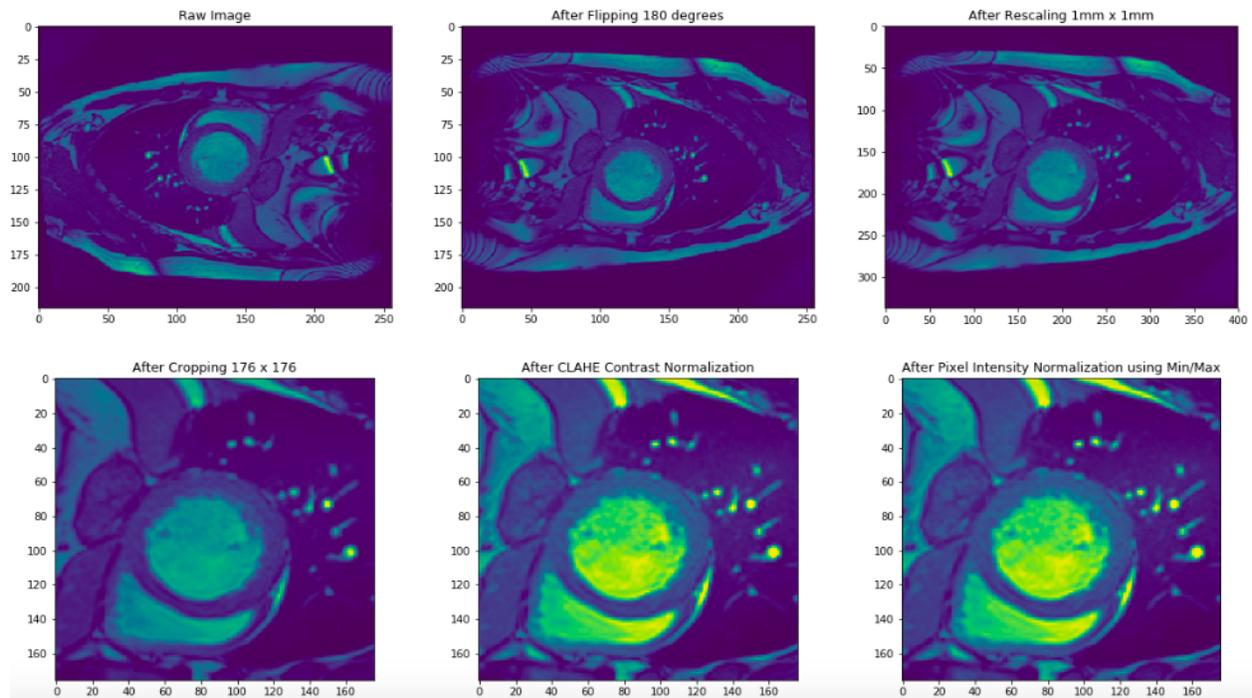

*Figure 8: ACDC Image Preprocessing Steps*

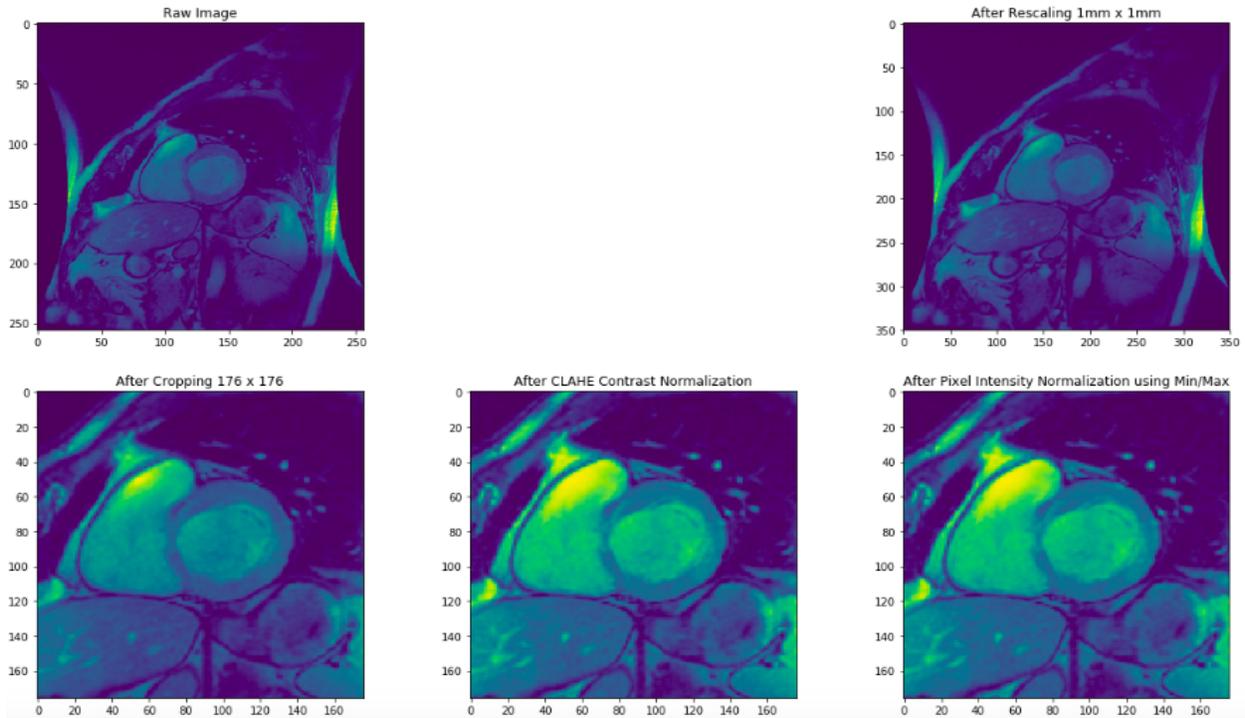
*Figure 9: Sunnybrook/MICCAI RV Image Preprocessing Steps*

Our results found that from frame to frame, there was not much variation, but slice to slice was very different at the first and last slice. From the other papers regarding segmentation over MRI, it was indicated that most of the Convolutional Neural Network(CNN) based techniques produce erroneous segmentation at basal and apical slices due to the following two major issues, namely:- (i) Uncertainties in the ground truth at valve level due to limited long-axis resolution of MRI, and (ii) Difficulty in exactly defining the apex and also presence of trabeculations near apex.

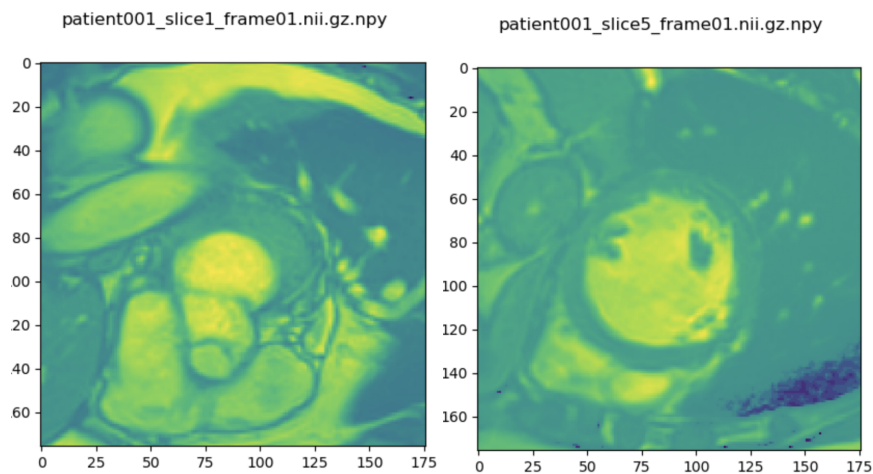
*Figure 10: An end slice (slice 1) compared to a center slice(slice 5)*

The pixel depth of MRI images in our dataset is 16 bits. However, the images have very low dynamic range with pixel intensities between 0 and 4000. Since applications include images with poor contrast due to glare or other reasons, normalization is done to bring the image, into a range that is more familiar or normal to the senses. Looking at Figures 4 & 5, it is evident that pixel intensity normalization will improve the segmentation results. From the images, we can also determine that we are able to preserve the ROI containing the LV/RV region even after images are cropped to 176x176 pixels. We chose 176x176 because other approaches we studied used a smaller image size and will be much faster to train the model. Eventually, we ended up applying the preprocessing steps for the papers we took inspiration from, Isensee for the 3D U-Net model we used and Khened for the 2D U-Net/DenseNet.

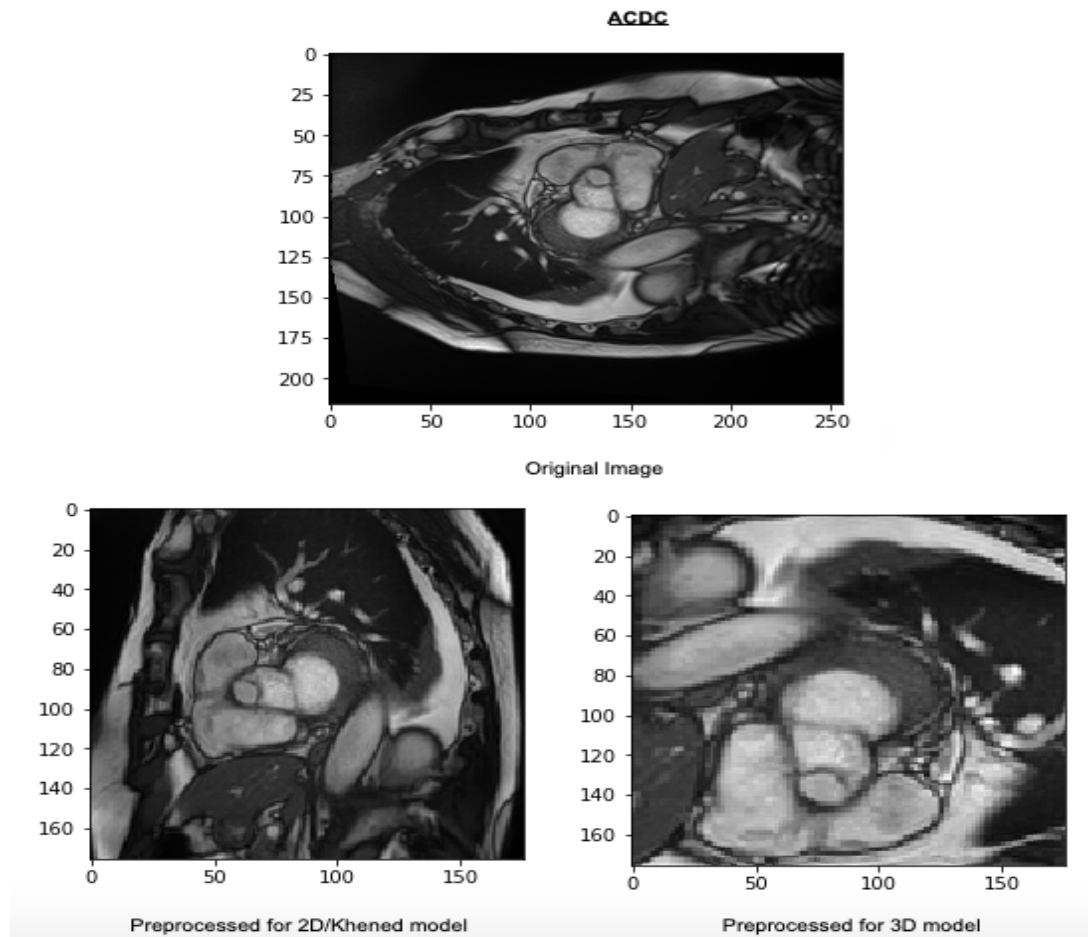

*Figure 11: Final Project Preprocessing*

**Models**

Our project's approach was to segment MRI results using the short-axis (SAX) slices. There are three models under consideration. The first is a 3D U-Net, as described by Isensee[1] to segment brain MRI images. We modified the Isensee 3-D code to work with 2-D data, and made changes to perform better with our cardiac data. Another model we are looking at is a DenseNet as described by Khened et al. [2], modified to work on the preprocessed data we created.

**2D U-Net**

We used U-Net, a deep learning model that was originally created for image segmentation in biomedical applications but has been successfully applied to other domains as well. The U-Net architecture combines low-level feature maps with high-level feature maps for precise pixel-level positioning. On each downsampling layer, the image size becomes 1/2 of the original, and the number of features becomes 2 times the original. On each upsampling layer, the image size becomes 2 times the original size, and the number of features is cut in half. In the upsampling operation, each output feature is merged with the features of the phase-contracted contraction network to complement the intermediate lost boundary information. The general U-Net architecture is shown in the figure below:

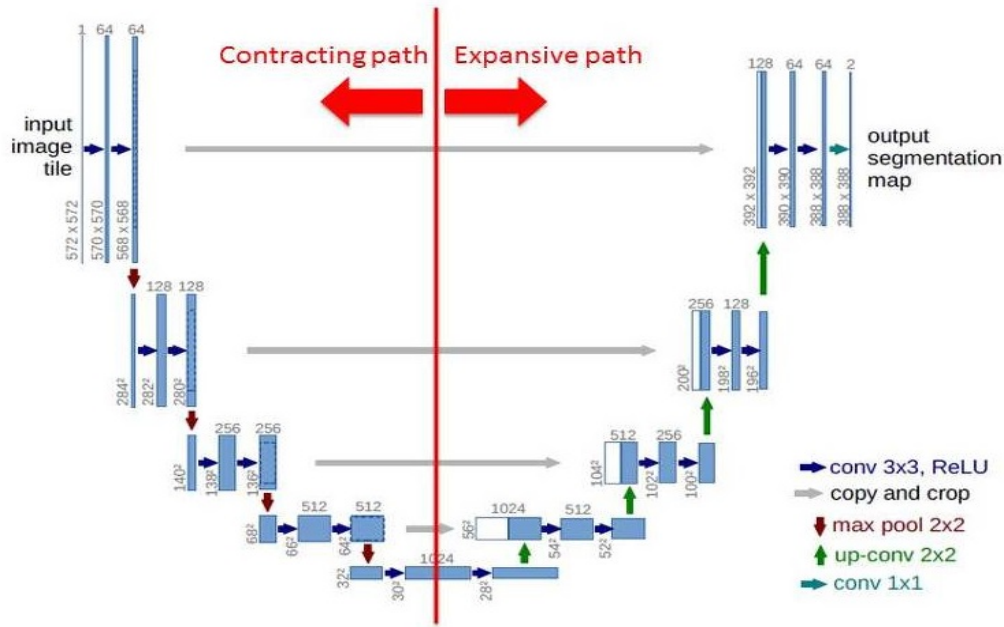

*Figure 12: General U-Net Architecture from [10]*

The U-Net model we used was derived from [1], and following the terminology there, the contracting path is referred to as the context pathway and the expansive path is referred to as the localization pathway. Each convolution block in the context pathway consists of a 2D convolution layer, a context module, and a 3x3 convolution with a stride of 2. Each context module is a pre-activation residual block [3] with a 3x3 convolution layer, a dropout layer, followed by another 3x3 convolution layer. Each convolution block in the localization pathway consists of an upscaling layer (size 2, stride 2), followed by a 3x3 convolution that halves the number of feature maps. Following the upsampling, feature maps from the localization pathway are concatenated with corresponding feature maps from the context pathway and passed to a localization module. A localization module consists of a 3x3 convolution followed by a 1x1 convolution and halves the number of feature maps. Segmentation layers at different levels are combined using element-wise summation to form the final network output. Our 2D U-Net has 2,770,825 parameters and is similar to the figure below, except that 2D filters instead of 3D filters are used.

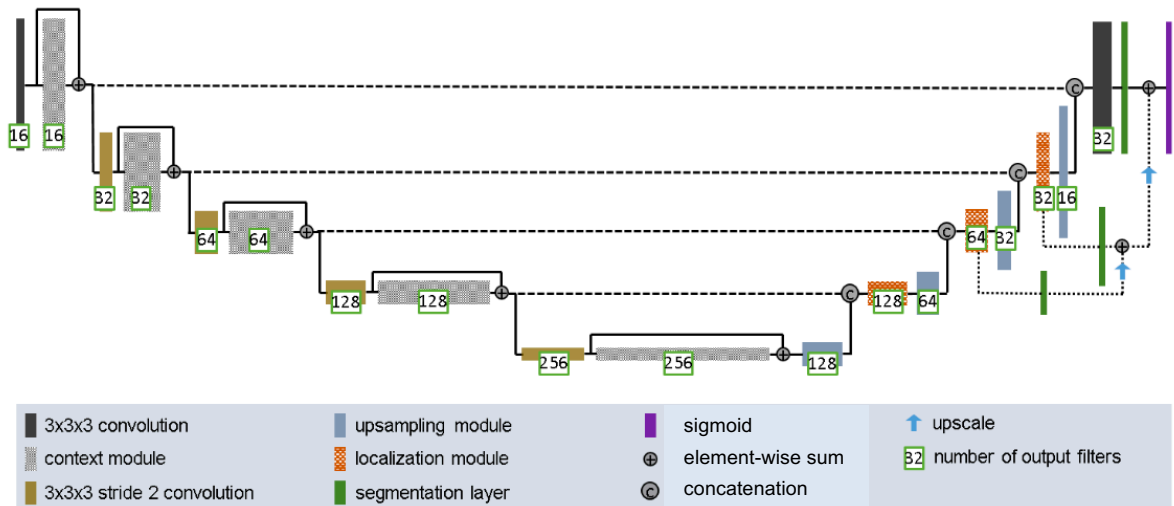

*Figure 13: 3D U-Net Model is based on [1]. 2D U-Net is the same except with 2D filters.*

**3D U-Net**

The difference between 3D U-Net model and 2D U-Net model is how the convolution is performed. 3D U-Net performs 3D convolution over x,y and z axis (i.e. over a cubic) while 2D U-Net works on only x and y (i.e. plane or single image). The 3D U-Net model we used has 33,049,417 parameters and is shown in the figure above. For ACDC dataset, under each patient, there are two frames which includes multiple slices (i.e. images) of each. 3D U-Net can perform 3D convolution over entire frame (i.e. all included slice images) at a time to learn context between slices when training while 2D U-Net runs 2D convolution on each slice one by one. For other datasets (namely, Sunnybrook and RV), raw image files are structured as slice by slice under each patient (i.e. each slice is a DICOM image file or NIfTI image file). So additional data preprocessing is required to combine all the slices belonging to the same frame together to turn into a single frame image file (i.e. a frame including 8-17 slices), then feed into pipeline for normal data preprocessing, training and predicting.

For the 3D U-Net model, the following image preprocessing/normalization steps were used:

- Center Cropping
    - 176 x 176 pixels
- Zero mean normalization
    - contrast normalization
- Bias field correction

Below is a visualization of a preprocessed MRI image and the model prediction compared to the ground truth:

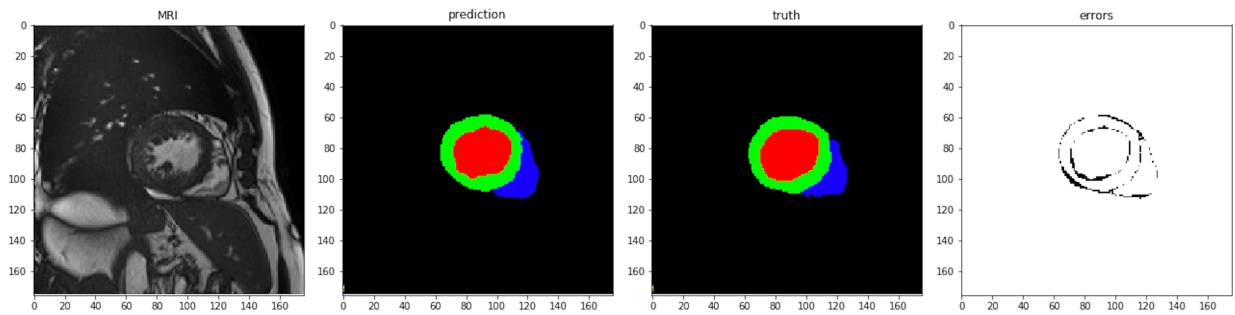

*Figure 14: 3D Model ACDC Single Image Output Results*

**DenseNet**

The Khened model utilizes a fully convolutional multi scale residual DenseNet for semantic segmentation. This architecture seen in the left figure below is comprised of a down sampling path as well as a upsampling path. The modular blocks that the architecture is comprised of are shown in more detail in the right figure below. From the Khened paper, "The input spatial resolution is recovered in the up sampling path by transposed convolutions, dense blocks and skip connections coming from the down sampling path"(10). For the up sampling operation, feature maps were added element wise with skip connections. Last, to generate the final label map of the segmentation, "the feature maps of the hindmost up-sampling component was convolved with a 1×1 convolution layer followed by a soft-max layer"(10).

For this model, segmentation of images was achieved by performing voxel wise classification where the output of the soft max layer gave the posterior probabilities for each class. The loss function used for training the network was a combination of cross entropy and dice loss. The network parameters were optimized to minimize both the loss functions in tandem. Additionally, an L2 weight decay penalty was added to the loss function as regularizer. The DenseNet model we used has 651,316 parameters and is shown in the figure below.

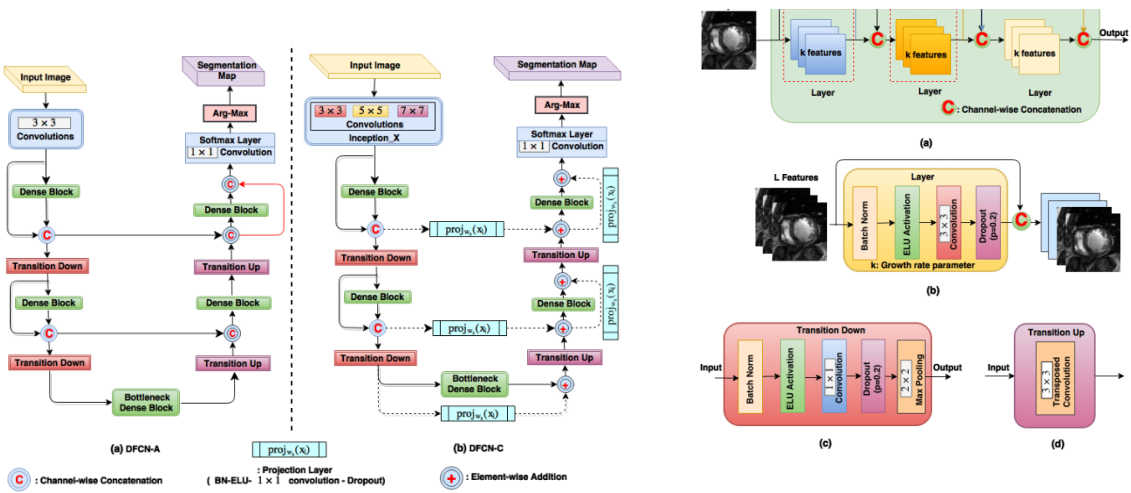

*Figure 15: DenseNet Overview*

For the DenseNet model the following image preprocessing steps were used:
- Orientation
    - ACDC images flipped to match format of other datasets
- Center Cropping
    - 176 x 176 pixels
- Z-Score
    - contrast normalization

Based on Khened's default model config settings, the following network hyper parameters were fixed and utilized by us:

● Number of max pooling operations(3)
● Growth rate of dense blocks(16)
● Max number of initial feature maps generated by the first convolution layers(3 times the growth rate)
● L2 weight decay(5e-6),
● Dropout rate(0.2)

**Data Augmentation**:

In order to be more generalizable, data augmentation was added to each model during training to better adapt to variations that might occur in other datasets. Data augmentation also allows the models to train on smaller datasets by simulating a larger dataset. Our current settings for augmentation have it alternating randomly by a coinflip for each image that gets passed through training so roughly 50% of all training images per epoch will be augmented. The following transformation parameters represent what augmentations are possible for each image(rotation,translation, zoom, flip):

> 'rotation_range': (-5, 5),
> 'translation_range_x': (-5, 5),
> 'translation_range_y': (-5, 5),
> 'zoom_range': (0.8, 1.2),
> 'do_flip': (True, True)

For each augmentation with ranges indicated, a certain number will be picked randomly for the image to be augmented by. For example, for rotation range, an image might be randomly chosen to rotate -2 degrees altering it counterclockwise. 'Do_flip' refers to horizontal/vertical flips of a given image.

**Metrics**

For our preprocessing process, we collect and record several metrics. The most important and most basic criteria is that our preprocessing does not lose data. To evaluate that, we record the number of patients and slices both before and after our preprocessing to confirm that both numbers are unchanged. Furthermore, we wanted to confirm that our preprocessing does what we expect it to. We crop to 176 X 176 and normalize the values. We collect before and after metrics on the image size and pixel values. As expected, all the files after preprocessing are 176 X 176, and all values are between 0 and 1.0.

The Dice similarity coefficient is used as a statistical validation metric to evaluate the performance of both the reproducibility of manual segmentations and the spatial overlap accuracy of automated probabilistic fractional segmentation of MR images.

Dice Coefficient metric is defined as

$$DC = \frac{2|A \cap B|}{|A| + |B|}$$

, where $A$ is the first set and $B$ is the second set of images. T From the equation, DC is always between 0 to 1. When it is closest to 1, the images are the most similar. As this fits well for our segmentation project, we use Dice Similarity Coefficient as one way to measure the performance of models.

Hausdorff distance is a symmetric measure of distance between two contours and is defined as:

$$H(P, G) = max(h(P, G), h(G, P))$$
$$h(P, G) = max_{pi \in P} \, min_{gi \in G} \, ||pi - gi||$$

Where P and G are the set of voxels enclosed by the predicted and ground truth contours delineating the object class in a medical volume respectively. A high Hausdorff value implies that the two contours do not closely match. The Hausdorff distance is computed in millimeter with spatial resolution and yet another metric we use to measure the performance of our predictions.

**Scalability**

Scalability requirements for cardiac image analysis project are defined as follows:
1. Model scalability
    a. Scalability in extending model prediction classes: The model should be easily scalable when the scope is extended to perform multi-class predictions.
    b. Scalability with image size: The model should work with different size of images without having to redesign the architecture.
    c. Scalability with pixel spacing: The model should work with different pixel spacing of images without having to redesign the architecture.
    d. Scalability with volume of training data: The model should meet acceptable performance even with limited training data.
    e. Scalability in adding datasets: Ensure that adding a new dataset is easy without modifying the models.
2. Scalability with execution environment
    a. CPU and GPU Execution environment: Ensure that the training and validation of the model can be done on either CPU or GPU environment.

b. Execution parallelism: It is required to run several experiments in parallel to evaluate the performance of the model with different preprocessing techniques and also to tune the hyper parameters of the model. The model should be able to be trained on multi GPUs.
3. Scalability with programming platform
   a. Scalability across multiple ML frameworks: Select a programming platform that provides abstraction for the model implementation. The platform should be able to run on various proven backend numerical computation libraries, machine learning libraries such as tensorFlow, Theano, CNTK etc.

The robustness requirements are captured in following areas :
1. Various backup options to save/restore the information
   a. Saving and restoring of execution environment (EC2 AMIs).
   b. Saving and restoring data (backups in S3, EBS volume snapshots).
2. Data storage formats optimized for specific data types
   a. The outputs of each stages are stored either using Image formats, or numpy arrays to minimize the processing overhead and storage memory.
3. Accuracy across datasets
   a. We removed any dataset specific features so that the models would get good results regardless of which dataset it is run against.

**Scalability with Image Size**

One of the key advantages that U-Net model has over regular convolution networks, is the scalability with image sizes. U-Net model does not have a Fully Connected Layer at the end. This makes the model flexible to work on images of different sizes. In other words, the model can be trained with different size images without redesigning the convolution layers inside the model. There is a requirement that the images be the same size, and that size has to be divisible by 2 a certain number of times, but that size can be changed by modifying a single configuration variable. Similarly, for the DenseNet model, image size can be changed as a config file property.

**Scalability with Volume of Training Data**

One of the challenges in medical image analysis is that very limited data is available for training. Conventional CNNs require large amount of training data to reach acceptable performance. Several studies have shown that U-Net model performs well even with limited training data. We will be using four different datasets: Automated Cardiac Diagnosis Challenge (ACDC), Sunnybrook (SB), LV 2011 Segmentation Challenge (LV) and MICCAI RV (RV) and verifying if increasing the volume of training data improves the model.

**Scalability in Extending Model for Multi-Class Predictions**

The scope of the our previous work [4] was to identify left ventricle in each SAX image, so the prediction task has only two classes (LV contour, background). However, the cardiac image analysis can be extended to locate other parts of heart such as right ventricle and myocardium, as in the case of ACDC. The goal of this project is to extend the model to handle multi-class predictions with very minimum changes to model design.

**Scalability across Datasets**

We are unlikely to ever get more data from the datasets we already use. To add more data, we would have to add another dataset. The datasets we already have show a lot of variation already. They are stored in different file formats, have different image sizes, different file spacing, different orientations, and have different ground truths. If the models weren't dataset agnostic, then adding a new dataset would require a lot of little changes in each of the three models. Instead, we developed a preprocessing step that would normalize the file size and spacing, and change the file format to numpy arrays. A new dataset would likely require updating that script to account for that dataset's idiosyncrasies, but updating that information once in a centralized location is better than letting each model try to handle it separately.

**Scalability with Execution Parallelism**

We are using Keras with Tensorflow with GPU support as our backend for high performance numerical computations. Using GPUs will speed up training the model. Using 4 GPUs increases the training speed by upto 75 times when compared to using a CPU. Keras can execute using multiple GPUs.

The GPU data parallelism works in the following manner:
1. First the batch is divided into sub batches (the input to the model) based on the number of GPUs available.
2. Model copy will be applied to each GPU.
3. The results are then aggregated as a single output using CPU.

The below considerations need to be kept in mind while executing multi-GPU [2]:
1. Speed does not scale linearly as the number of GPUs increases, as synchronization between GPUs is required.
2. Batch size will be further divided based on the number of GPUs available. Hence this needs to be carefully chosen as it affects the model weights and each sub-batch needs to fit the memory available on each GPU.
3. It is important to feed all the GPUs with data. It can happen that the very last batch of the epoch has less data than defined (because the size of your dataset can not be divided exactly by the size of your batch). This might cause some GPUs not to receive any data during the last step.
4. The *allow_growth* and *per_process_gpu_memory_fraction* Tensorflow options are strongly recommended for configuring when multiple jobs occupy the same GPU device. These two options are only available when running the jobs in multi-gpu mode.
5. *CUDA_VISIBLE_DEVICES* environment variable is set to make the GPU devices available. In addition, the gpus parameter passed into multi_gpu_model in keras has to be equal or less than the visible GPU devices specified.
6. At model and weights saving steps, the template model should be referenced rather than the model returned by multi_gpu_model function.

Our pipeline allows the same model to be executed on single GPU and multi GPUs instances. In our experience, we saw that using 4 GPUs decreases the training time by up to 3.5 times compared to using a single GPU.

In terms of our model performance training methods, we trained the 2D U-Net and DenseNet models trained on a single GPU (1 Tesla V100 and 16 GB memory) instance. For 3D U-Net, we trained on 4 GPUS (4 Tesla V100 and 64 GB memory) instance due to the extra computational power needed.

In terms of each models time taken to train for the experiments we ran: 2D U-Net took 3 to 58 minutes, Khened DenseNet took 20 to 185 minutes, and 3D U-Net took 25 minutes to 7 hours 45 minutes.

**Scalability across Multiple ML Frameworks**

Keras is a high-level neural network library, written in Python. It is capable of running on top of several Machine Learning and numerical computation libraries such as TensorFlow, PyTorch, CNTK, or Theano. Keras hides the interface details of each of these backend libraries through abstraction. This will enable users to switch the backends without changing the implementation. For this project we have experimented only with tensorflow backend. Keras provide rich documentation about how to change the backend in the following document https://keras.io/backend/.

**Dataset Experiments**

Going off the below experiment tables, we carried out the appropriate training/predictions for each of the three models, DenseNet, 2D U-Net, and 3D U-Net. This was done utilizing Khened preprocessing for the 2D U-Net and Khened models whereas Isensee preprocessing technique will be used by 3D U-Net model. Additionally, both preprocessing techniques will be cropped the images to 176 x 176 pixels.

These below figures denote what experiments we carried out on the model regarding what datasets we trained and predicted on which changed from our last reported plan.

| Experiment Number | 1a | 1b | 1c | 1d | 2a | 2b | 2c |
|---|---|---|---|---|---|---|---|
| Training Data | ACDC | LV | RV | SB | ACDC | ACDC | ACDC |
| Test Data | ACDC | LV | RV | SB | LV | RV | SB |

| Experiment Number | 3.1a | 3.1b | 3.1c | 3.2a | 3.2b | 3.2c |
|---|---|---|---|---|---|---|
| Training Data | ACDC LV | ACDC RV | ACDC SB | ACDC LV | ACDC RV | ACDC SB |

| Test Data | ACDC | ACDC | ACDC | LV | RV | SB |

| Experiment Number | 4a | 4b | 4c |
|---|---|---|---|
| Training Data | ACDC<br>LV<br>SB | ACDC<br>LV<br>SB | ACDC<br>LV<br>SB |
| Test Data | ACDC | LV | SB |

*Figure 16: Experiments Summary*

The plots below indicate what models contained the highest scoring records for certain metrics/datasets to help give an idea where each model performs best:

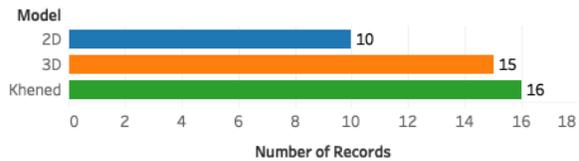
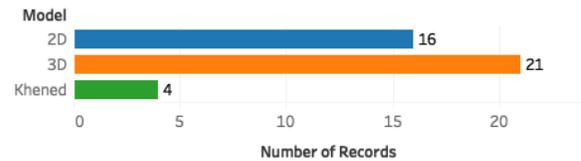
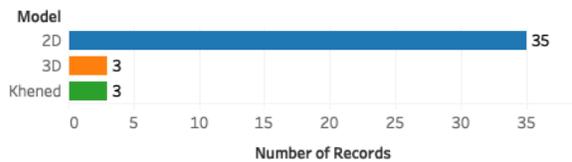
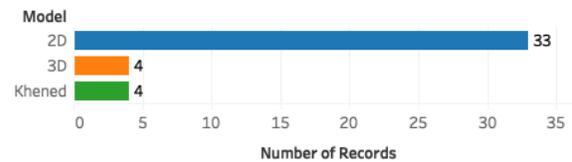

*Figure 17: Overall model results with number of top placings*

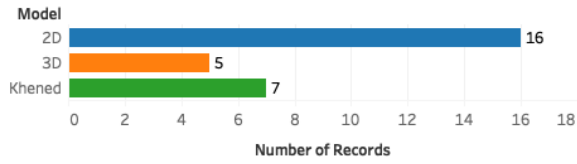
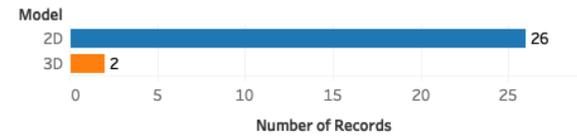
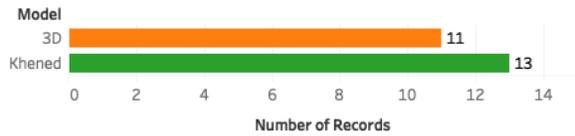
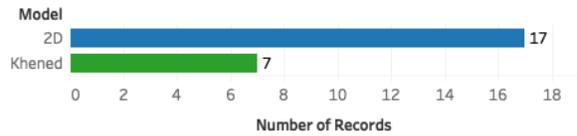
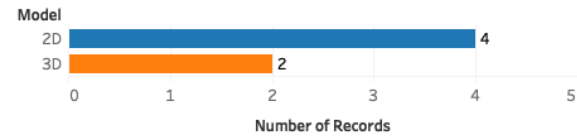
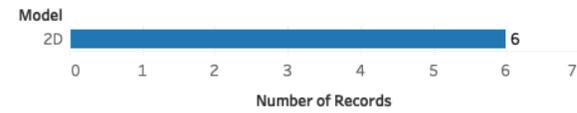
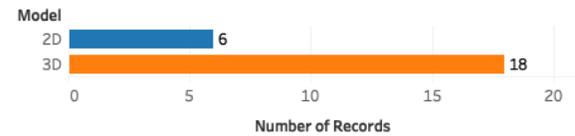
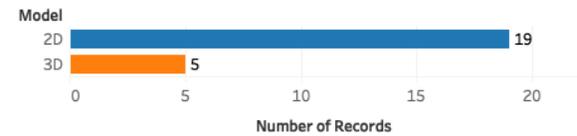

*Figure 18: Model Results separated by dataset*

Based on the results below, it's clear to see how generalized the various models are. It can be seen that 2D model loses the least performance when trained on ACDC compared to being trained on the same dataset as the test set.

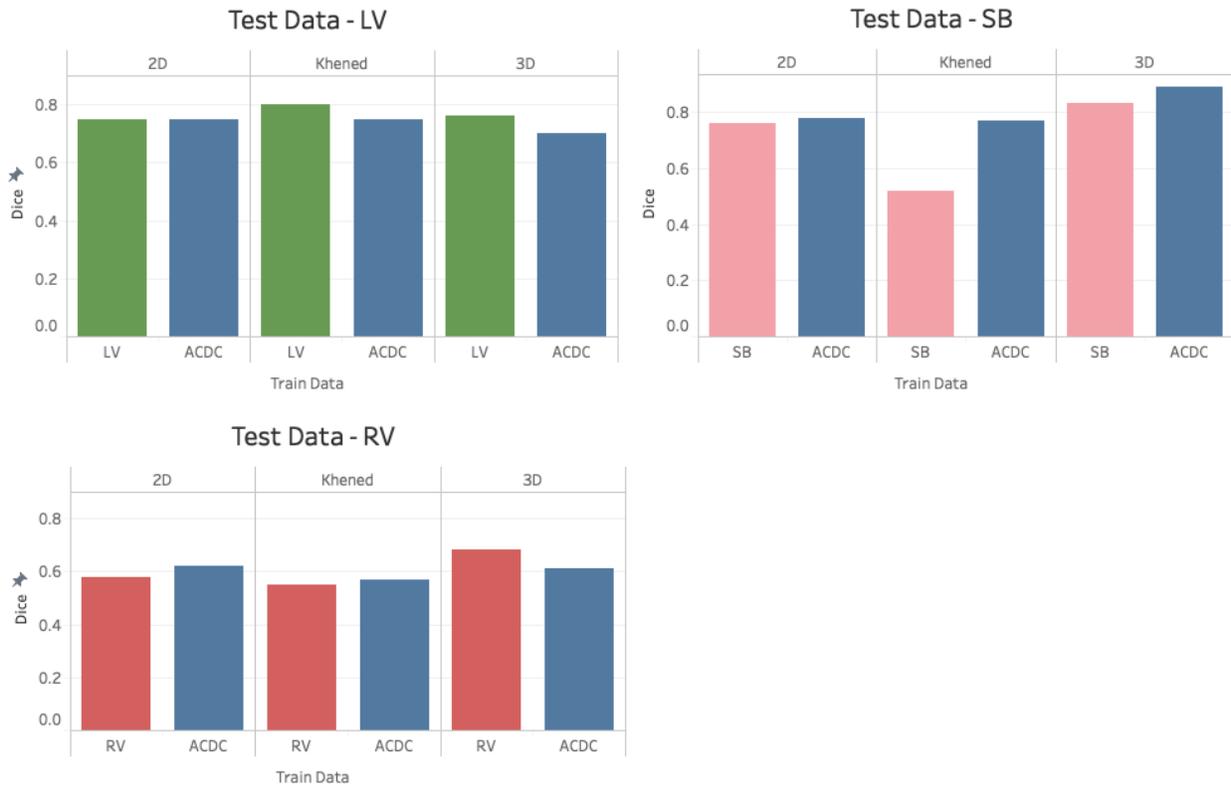

*Figure 19: Model generalization experiment results*

The other experiments aimed to determine the impact of training data volume on the model performance. We observed that in general adding more data results in a better trained model.

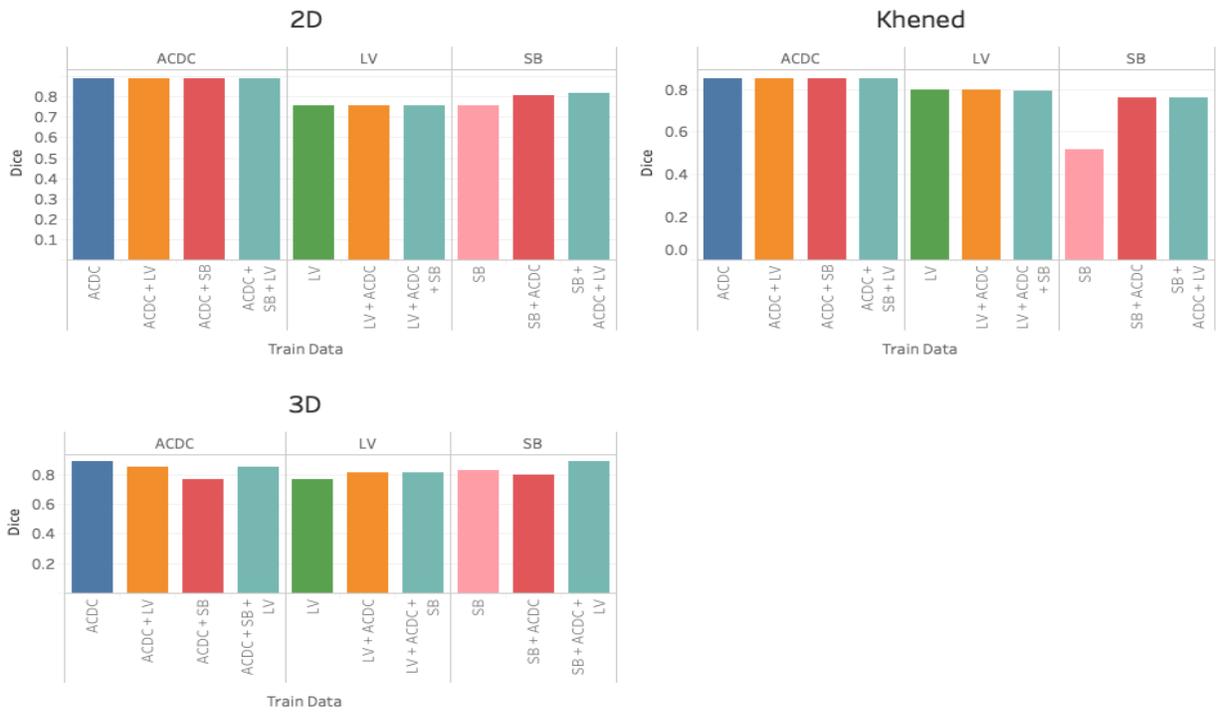

*Figure 20: Left Ventricle Dataset Size Improvements Experiments*

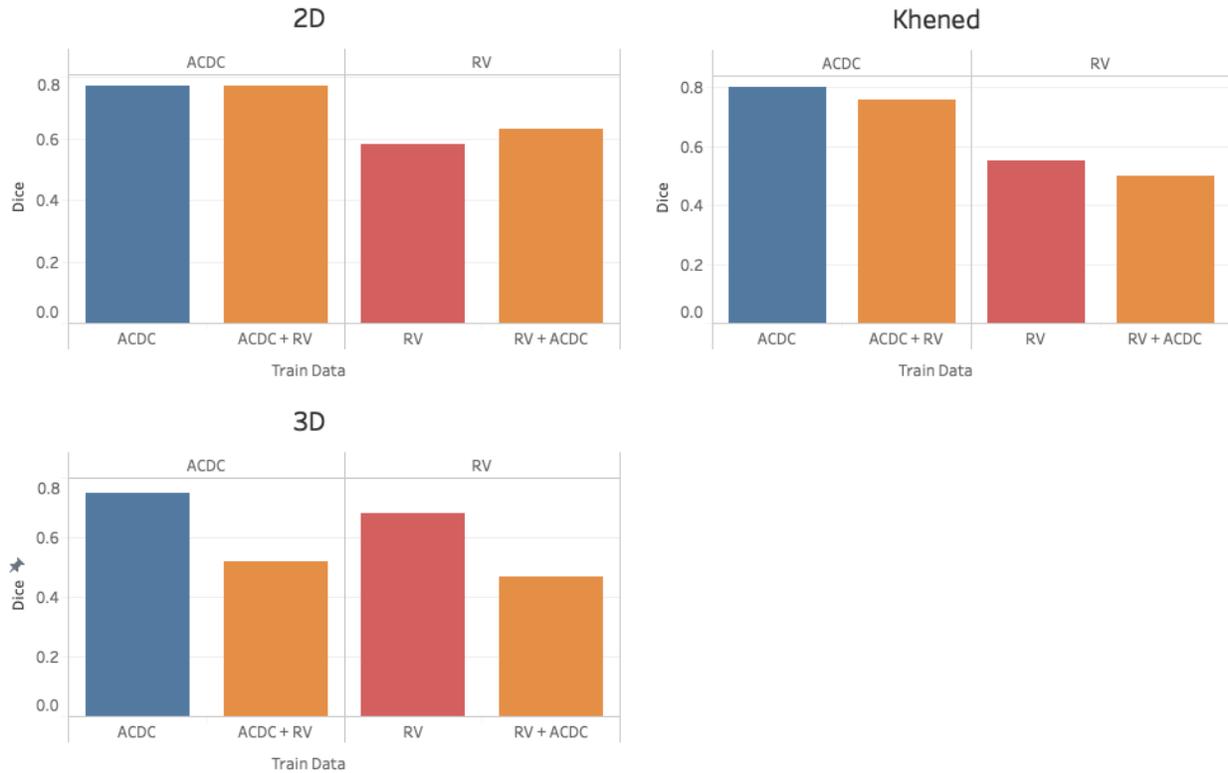

*Figure 21: Right Ventricle Dataset Size Improvements Experiments*

The 2D model performed better on RV data when both ACDC and RV data was used to trained the model. But that was not true for the other two models.

Training the 2D and Khened models took relatively less time when compared to training the 3D model. The 2D and Khened models were trained on single GPU EC2 instance (p3.2xlarge) which consists of 1 Tesla V100 GPU with 16GB memory. Since it takes a very long time to train the 3D model on this instance, more than a day when combining the datasets, we used 4 GPUs EC2 instance (p3.8xlarge) to train the 3D model. This instance has 4 Tesla V100 GPUs with 64GB memory. The training times for the various models are shown in the figure.

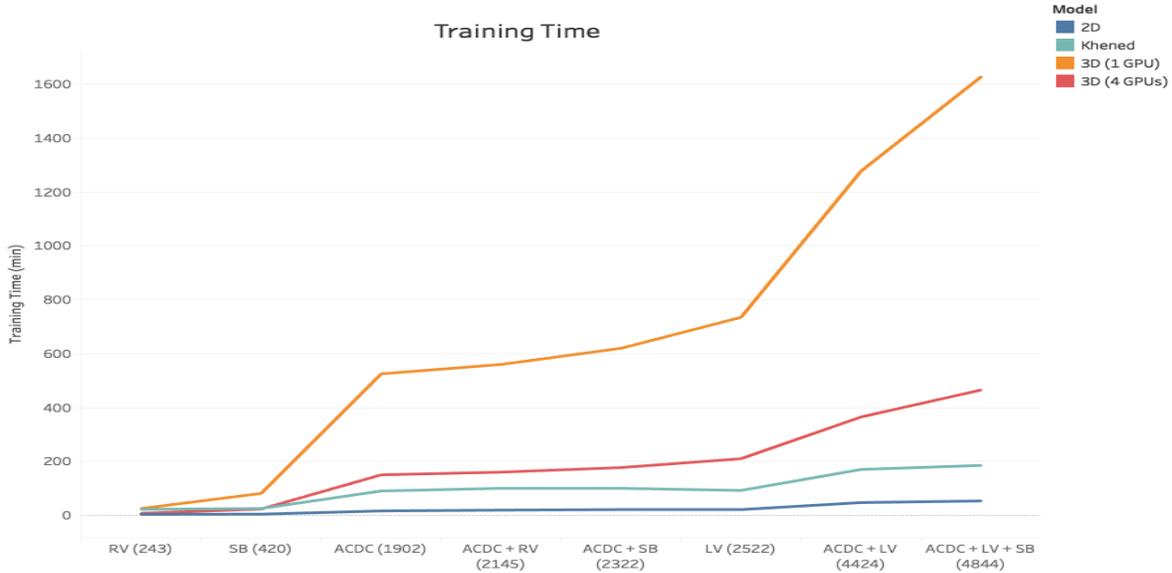

*Figure 22: Model Training Times across dataset experiments*

**Post-Processing**

Three post processing steps were tested against the results of the 2D U-Net. The first was to find any images that had multiple Left Ventricle Cavity (LVC) segments, and only keep the largest segment. There were not many predictions that had multiple LVC segments, but it improved the Dice score by about 1%. This treatment was rejected for Left Ventricle Myocardium (LVM) segments because they are very thin, so breaks in continuity were not uncommon. For the RV, we eliminated any predictions that were too far away from the LVC or LVM. This did not help our Dice score, but significantly improved our Hausdorff distances. This is not unexpected, since there were very few images this applied to, and the images it did apply to had unusual RVs. These images had very little overlap, so modifying the denominator did not affect the Dice score very much. However, the Hausdorff distance is much more sensitive to bad predictions, as it is a measure of how far away the worst predicted pixel is, making deleting bad predictions an effective way to raise the score. There is one experiment where the Dice score was significantly impacted: The train on ACDC + RV and predict on RV experiment dropped precipitously from .63 to .41, indicating that some good predictions are being deleted. We believe this is due to the RV dataset not having labels for the LV, which means the LV predictions are affected negatively. Since we delete RV predictions based on their proximity to the LV predictions, bad LV predictions will make this post processing step undesirable. The final step of postprocessing is to take any small islands of predicted 0, and filling them in with the nearest neighbor. We noticed that for the 2D U-Net, it would often predict a ring of 0s in between the LVM and LVC. While we do not know which it is, the Dice score is calculated in such a way that guessing with 50% accuracy will help the Dice score. This did improve the Dice score, but it is an effect lower than 1%, and is often lost to rounding in our results.

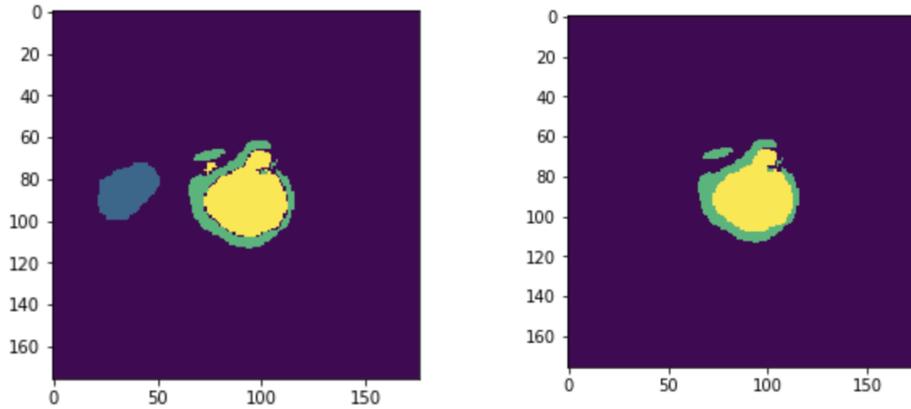

*Figure 23: Above: An image showing all 3 post processing steps. There is a yellow (LVC) island that is deleted as the main LVC is larger. The blue RV is deleted as it is too far away from the LV. The gaps between the LVC and LVM are filled in. Also of note is that the green LVM island is not deleted, as the LVM is not subjected to the rule that deletes islands.*

**Findings**

The first finding we had was that it is possible to get good results for segmenting the left and right ventricle using any of 3D U-Net, 2D U-Net or DenseNet. When trained and tested on ACDC, all the dice scores were good. The models had some variation, but all of the scores for the inner surface of the LV were in the low 90s, and the more irregular outer wall of the LV and the RV had scores in the high 70s or low 80s. Running through the different experiments showed some more distinctions between the models. For example, DenseNet requires the most, best data. When it trained on ACDC, it had the best score for the right ventricle, but when it trained on MICCAI RV, a very small dataset, it had the worst score. Similarly, Sunnybrook is a midsize dataset with relatively low image quality, and DenseNet had the worst performance when training on Sunnybrook alone. Training on ACDC and Sunnybrook together helped every model compared to Sunnybrook alone, but the effect was the most pronounced in DenseNet. The LV Segmentation challenge is a large dataset with high quality images, and DenseNet has the best score in every category. 2D U-Net shows the opposite characteristics. It does not win very many categories where the training and test dataset are the same, but does do very well when the training dataset is a combination of multiple datasets, suggesting that it is very generalizable. Ideally, a hospital or other consumer will be able to predict without training a model on their own data, and an ability to train and test on different datasets without degrading performance is important to that goal. 3D U-Net has the most overall dice score wins. The 3D U-Net uses the context of the slice's position and neighbors to improve the prediction, but it takes the most dataset specific work, and the most training time.

The findings can be summarized as follows:

- 2D
    - Generalizes well
    - Adding more data samples helps
    - Better segmentation results for LV dataset
- Khened

- - Does not perform well on SB dataset (dark images)
    - Performance decreases if trained and tested on different datasets
    - Adding more data samples helps for SB, but not RV dataset
    - ED phase predictions better than ES
- 3D
    - Have to be able to align the different slices for each frame to get correct results
    - Adding more data samples helps for LV, but not for ACDC
    - Doesn't generalize well
    - Takes a very long time to train on a single GPU (upto a day)

**References**

1. Isensee, Fabian, et al. "Brain Tumor Segmentation and Radiomics Survival Prediction: Contribution to the BRATS 2017 Challenge." ArXiv.org, Cornell University, 28 Feb. 2018, arxiv.org/abs/1802.10508.
2. Khened, Mahendra, et al. "Fully Convolutional Multi-Scale Residual DenseNets for Cardiac Segmentation and Automated Cardiac Diagnosis Using Ensemble of Classifiers." Medical Image Analysis, Elsevier, 19 Oct. 2018, www.sciencedirect.com/science/article/pii/S136184151830848X.
3. K. He, X. Zhang, S. Ren, and J. Sun, "Identity mappings in deep residual net-works," inECCV. Springer, 2016, pp. 630–645.
4. E. Abdelmaguid, J. Huang, S. Kenchareddy, D. Singla, L. Wilke, M. H. Nguyen, and I. Altintas, "Left ventricle seg- mentation and volume estimation on cardiac mri using deep learning," in eprint arXiv:1809.06247 [cs.CV], 2018. https://arxiv.org/ftp/arxiv/papers/1809/1809.06247.pdf
5. https://lmb.informatik.uni-freiburg.de/people/ronneber/u-net/
6. Cardiac CINE MRI process: https://www.youtube.com/watch?v=yZXfvXiJwLc
7. Cardiac MRI tutorial: http://www.vhlab.umn.edu/atlas/cardiac-mri-tutorial/index.shtml
8. Machine Learning Blog & Software Development News: http://blog.datumbox.com/5-tips-for-multi-gpu-training-with-keras/
9. 3D U-Net Convolution Neural Network with Keras github: https://github.com/ellisdg/3DUnetCNN
10. U-Net: Convolutional Networks for Biomedical Image Segmentation: https://arxiv.org/pdf/1505.04597.pdf
11. Fully Convolutional Multi-scale Residual DenseNets for Cardiac Segmentation and Automated Cardiac Diagnosis using Ensemble of Classifiers : https://arxiv.org/abs/1801.05173
12. Automatic Cardiac Disease Assessment on cine-MRI via Time-Series Segmentation and Domain Specific Features https://arxiv.org/abs/1707.00587